\documentclass[11pt]{article}
\usepackage{authblk}
\usepackage{enumitem}
\usepackage{longtable}
\usepackage{a4wide}
\usepackage[numbers]{natbib}
\usepackage[hidelinks]{hyperref}
\usepackage{amsmath}
\usepackage{amsfonts}
\usepackage{setspace}
\usepackage{graphicx}
\usepackage{caption}
\usepackage{subcaption}

\spacing{1.5}
\nonumber

\title{Multidimensional well-being of US households at a fine spatial scale using fused household surveys: fusionACS}

\author[1]{Kevin Ummel}
\author[2,3*]{Miguel Poblete-Cazenave}
\author[4]{Karthik Akkiraju}
\author[5]{Nick Graetz}
\author[1]{Hero Ashman}
\author[1]{Cora Kingdon}
\author[1]{Steven Herrera Tenorio}
\author[1]{Aaryaman ``Sunny'' Singhal}
\author[1]{Daniel Aldana Cohen}
\author[4,3]{Narasimha D. Rao}
\affil[1]{University of California, Berkeley, Socio-Spatial Climate Collaborative, Berkeley, 94720, USA}
\affil[2]{VU Amsterdam, Institute for Environmental Studies (IVM), Amsterdam, 1081, Netherlands}
\affil[3]{International Institute for Applied Systems Analysis (IIASA), Energy, Climate, and Environment Program, Laxenburg, A-2361, Austria}
\affil[4]{Yale University, School of the Environment, New Haven, 06511, USA}
\affil[5]{Princeton University, Princeton, 08544, USA}

\affil[*]{Corresponding author(s): Miguel Poblete-Cazenave (m.a.poblete.cazenave@vu.nl)}

\date{}

\begin{document} 

\maketitle

\abstract{Social science often relies on surveys of households and individuals. Dozens of such surveys are regularly administered by the U.S. government. However, they field independent, unconnected samples with specialized questions, limiting research questions to those that can be answered by a single survey. The fusionACS project seeks to integrate data from multiple U.S. household surveys by statistically "fusing" variables from "donor" surveys onto American Community Survey (ACS) microdata. This results in an integrated microdataset of household attributes and well-being dimensions that can be analyzed to address research questions  in ways that are not currently possible. The presented data comprise the fusion onto the ACS of select donor variables from the Residential Energy Consumption Survey (RECS) of 2015, the National Household Transportation Survey (NHTS) of 2017, the American Housing Survey (AHS) of 2019, and the Consumer Expenditure Survey - Interview (CEI) for the years 2015-2019. The underlying statistical techniques are included in an open-source $R$ package, fusionModel, that provides generic tools for the creation, analysis, and validation of fused microdata.}

\section*{Background \& Summary}

Ideally, the research community would have access to a “comprehensive survey” that employs a large sample size, asks many questions on various topics, is representative of the general population, and enjoys perfect recall and accuracy. Such a survey would allow researchers to examine spatial patterns at higher resolution, analyze differences across detailed population subgroups, explore relationships among a wide range of phenomena, and build detailed micro-simulation models to anticipate policy impacts across households and communities. Unfortunately, a truly comprehensive survey is impossible. Budgets and sample sizes are limited; respondent participation suffers if too many questions are asked; and the scope of social phenomena is too large for a single survey instrument. In practice, a diverse collection of surveys exists at any one time, varying in size, subject matter, structure, and provenance.

Practitioners regularly impute or otherwise predict a variable or two from one dataset on to another. Piecemeal, $ad$ $hoc$ data fusion is a common necessity of quantitative research. Proper data fusion, on the other hand, seeks to systematically integrate two different samples into one microdata set. The desire to “fuse” or otherwise integrate independent datasets has a long history, dating to at least the early 1970’s\citep{Ruggles1974,Alter1974}. The most prominent examples of data fusion have involved administrative record linkage\citep{Chetty2014,Chetty2016,Chetty2020,Meyer2019,Medalia2019}. This consists of exact matching or probabilistic linking of independent datasets, using observable information like social security numbers, names, or birth dates of individuals. Record linkage, the gold standard, can yield important insights and high levels of statistical confidence. However, it is rarely feasible for the kinds of publicly available microdata that most researchers use day-to-day (nevermind the difficulty of accessing administrative data). 

The principal aim of the fusionACS project is to maximize the amount of information that can be extracted from the existing array of U.S. social surveys. This is accomplished through statistical “fusion” of disparate surveys in an attempt to simulate a more comprehensive survey. The technique uses the American Community Survey (ACS) – the largest U.S. household survey – as the “data backbone” of this process. Variables in “donor” surveys are fused onto ACS Public Use Microdata Sample (PUMS) microdata to produce simulated values for variables unique to the donor. This results in probabilistic estimates of how ACS respondents might have answered the donor survey questionnaire. Respondent characteristics that are common to both the donor and the ACS (e.g. income) – as well as spatial information that can be merged to both (e.g. characteristics of the local built environment) – are used to model donor variable outcomes using machine learning techniques.

The fusionACS pipeline produces ACS PUMS microdata with donor survey variables fused (simulated) for each respondent household ($\sim$1.3 million per year). This output can be used to perform any kind of analysis typically applied to microdata, with the added benefit that analyses can use variables from both the ACS and donor survey questionnaires. Moreover, the output microdata can be used to produce estimates for specific locales at the level of individual Public Use Microdata Areas ($\sim$2,300 nationwide), a higher level of granularity than that available in most donor surveys. Additionally, by passing the microdata through an additional spatial downscaling step\citep{Graetz2022}, estimates can be produced for areas as small as individual block groups. 

\begin{figure}[ht]
\centering
\includegraphics[width=\linewidth]{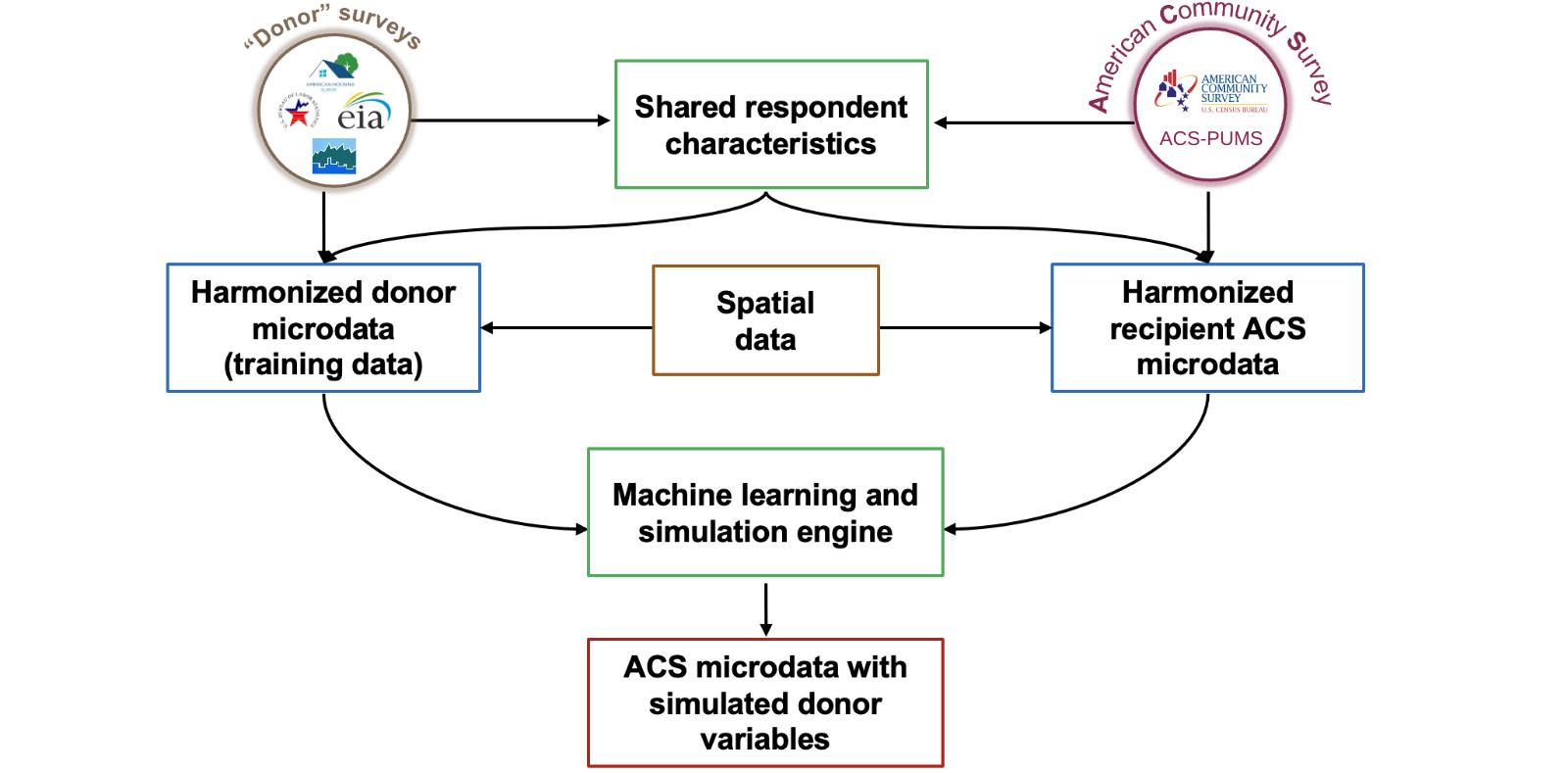}
\caption{fusionACS: Flow of data}
\label{fig:fig_1}
\end{figure}

The fusionACS “platform” consists of two packages written in the $R$ programming language. The fusionModel package provides an open-source interface for general data fusion (i.e. modeling and analytical tools). A separate, data processing package (fusionData) is used to generate the data inputs needed to fuse variables from a range of U.S. social surveys onto ACS microdata.  For a given candidate donor survey, the data processing and analytical “pipeline” consists of the following steps (see  Figure \ref{fig:fig_1}):
\begin{enumerate}[noitemsep]
\item Ingest raw survey data to produce standardized microdata and documentation.
\item Harmonize variables in the donor survey with conceptually-similar variables in the ACS.
\item Prepare clean, structured, and consistent donor and ACS microdata.
\item Train machine learning models on the donor microdata.
\item Fuse the donor’s unique variables to ACS microdata.
\item Validate the fused microdata to gauge the quality of the fusion process.
\item Analyze the fused microdata to calculate estimates and margins of error.
\end{enumerate}

Steps 1-3 are part of the fusionData package. Steps 4-7 are carried out using the fusionModel package. 

In principle, any survey of U.S. households or individuals circa 2005 or later is a candidate for fusion. Ideal donor surveys are those with larger sample sizes, respondent characteristics that overlap with ACS variables, and more detailed information on respondent location. Absence of these factors does not preclude usage but will affect the associated uncertainty.

\section*{Methods}

In the context of fusionACS, we are interested in the following problem:

\leftskip=0.2in
\rightskip=0.2in
\textit{
We have microdata from two independent surveys, $A$ and $B$, that sample the same underlying population and time period (e.g. occupied U.S. households nationwide in 2018). We specify that $A$ is the “recipient” dataset and $B$ is the “donor”. Survey $A$ is the American Community Survey and invariably has a larger sample size than $B$ ($N_a > N_b$). The goal is to generate a new dataset, $C$, that has the original survey responses of $A$ plus a realistic representation of how each respondent in $A$ might have answered the questionnaire of $B$. To do this, we identify a set of “harmonized” variables, $X$, that are common to both surveys; in practice, these are often things like household size, income, respondent age, race, etc. We then fuse a set of variables unique to $B$ – call them $Z$, the “fusion variables” – onto the original microdata of $A$, conditional on $X$.
}

\leftskip=0in
\rightskip=0in
This has generally been posed as a “statistical matching” problem\citep{DOrazio2006} whereby records from the donor microdata ($B$) are matched to a statistically-similar record in the recipient ($A$). Variables common to both datasets ($X$) are used to calculate similarity between records. For each record in $A$, a set of similar records are identified in $B$; e.g. using a $k$-nearest neighbor algorithm. A single record in $B$ is selected from this set and the variables unique to the donor ($Z$) are added (fused) to the matched record in $A$. A “mixed method” variant of this approach (see, e.g., Section 3.1.3 in Lewaa et al.\citep{Lewaa2021}) fits statistical models to $B$ to estimate the conditional expectation of $Z|X$. The models are used to predict $Z|X$ for both $A$ and $B$ ($Z\times a$ and $Z\times b$, respectively), possibly adding a random residual. The similarity of donor and recipient records is then calculated using $Z\times a$ and $Z\times b$ (rather than $X$) and the ultimate fusion of $Z$ proceeds as in the statistical matching case. The mixed method is effectively an implementation of predictive mean matching (PMM) first developed by Rubin\citep{Rubin1986} in the context of statistical matching and then extended to missing data imputation by Little\citep{Little1988}. Mixed, PMM-based techniques offer a number of advantages, including some protection against model misspecification (in the stochastic case) and a more defensible (and fast) calculation of record similarity, since it avoids calculating similarity across $X$ variables of possibly mixed types and varying levels of relevance in explaining $Z$.

Statistical matching techniques – mixed or otherwise – generally fuse complete records from the donor. This is a practical advantage, since it ensures that multivariate relationships among the fused variables are not obviously erroneous. But complete matching also introduces the possibility that donor observations will be repeated – possibly many times – in the fused dataset, increasing the risk that real-world variance is under-represented in the fused dataset. Intuitively, matching of complete records is most sensible when the donor’s sample size is at least as large as the recipient’s ($N_b \geq N_a$) and the number of variables to be fused is small. Neither condition holds for fusionACS use cases. A useful variant comes from the imputation literature\citep{Little2019}, where the insertion of complete records is impossible due to the typical sparsity of missing data. Imputation techniques usually proceed sequentially, filling in missing values one variable at a time or, alternatively, by sequential “blocks” of variables that are imputed jointly (e.g. see the popular mice imputation package\citep{Van2011}). A related literature in the area of data synthesis for statistical disclosure control\citep{Drechsler2011} also relies on sequential (“chained”) generation of synthetic variables. For example, Reiter\citep{Reiter2005} introduced the use of machine learning decision trees\citep{Breiman1984} to create wholly synthetic versions of survey microdata that do not rely on record matching\citep{Nowok2016,Bowen2020}. However, the goal in these cases is the synthesis of a single dataset for purposes of disclosure control, not the fusion of separate datasets.

The fusion strategy implemented in the fusionModel package borrows and expands upon ideas from the statistical matching\citep{DOrazio2006}, imputation\citep{Little2019}, and data synthesis\citep{Drechsler2011} literature to create a flexible data fusion tool. It employs variable-$k$, conditional expectation matching that leverages high-performance gradient boosting algorithms. The methodology and code is tailored for intended fusionACS applications, allowing fusion of many variables, individually or in blocks, and efficient computation when the recipient (the ACS in the case of fusionACS) is large relative to the donor.
Specifically, the goal was to create a data fusion tool that meets the following requirements:
\begin{itemize}[noitemsep]
\item Accommodate donor and recipient datasets with divergent sample sizes
\item Handle continuous, categorical, and semi-continuous (zero-inflated) variable types
\item Ensure realistic values for fused variables
\item Scale efficiently for larger datasets
\item Fuse variables “one-by-one” or in “blocks”
\item Employ a data modeling approach that:
\vspace{-0.1cm}
\begin{itemize}[noitemsep]
\item Makes no distributional assumptions (i.e. non-parametric)
\item Automatically detects non-linear and interaction effects
\item Automatically selects predictor variables from a potentially large set
\item Ability to prevent overfitting (e.g. cross-validation)
\end{itemize}
\end{itemize}

There are practical limits to this process, generally reflected in declining confidence in results as more is asked of the underlying data. For this reason, uncertainty estimation (via multiple implicates and associated analytical tools) is an important part of fusionACS’s development. Ideally, researchers are able to ask any question of fusion output and then decide if the answer’s associated uncertainty is suitable for the intended analysis.

\subsection*{General Strategy}

Consider the simple case where we fuse a single, categorical variable $Z$ consisting of $v$ classes. Using the notation from above, we fit a model to the donor data, $G = f(Z|X_b)$. $G$ is used to predict conditional expectations for each recipient observation, $D_a = G(X_a)$. In this case, $D_a$ is a $N_a \times v$ matrix of conditional probabilities from which $N_a$ simulated class outcomes ($Z_a$) are probabilistically drawn. The statistical model, $G$, consists of a LightGBM\citep{Ke2017} gradient boosting model that minimizes the cross-validated log-loss. The categorical case is comparatively straightforward and easily implemented.

Now consider fusing a single, positive continuous variable $Z$. In this case, we use multiple models to estimate the conditional distribution of $Z|X$. Let $G_u = f(Z_u|X_b)$ estimate the conditional mean and $G_q = f(Z_q|X_b)$ estimate conditional quantiles ($q$) associated with $p$ equally-spaced percentiles. This yields $p + 1$ cross-validated LightGBM models. $G_u$ minimizes the cross-validated squared error (L2) loss; $G_q$ minimizes the cross-validated quantile (pinball) loss. Training models for large $p$ is expensive; by default, we use $p = 3$ with percentiles $\{0.166, 0.5, 0.833\}$. The conditional expectations of the recipient observations, $D_a$, consists of a $N_a \times (p + 1)$ matrix of conditional mean and quantiles.

Unlike in the categorical case, there is no obvious way to simulate $Z_a$ from $D_a$. Common parametric assumptions are not ideal, since the conditional expectations imply unknown and (quite often) decidedly non-normal distributions. One option is to extend PMM to the current context, resulting in generalized “conditional expectation matching”. In this case, we derive $D_b$ by predicting $G_u$ and $G_q$ back onto the original training data, then find the $k$ nearest neighbors ($k$NN) in $D_b$ associated with each observation in $D_a$. This is analogous to conventional PMM, except that we use Euclidean distance based on $p + 1$ conditional expectations to find the nearest neighbors. Each $Z_a$ is then sampled randomly from the $k$ nearest neighbors in the donor.

There are drawbacks to this approach. First, it fundamentally differs from that used for a single categorical variable. In the categorical case, $D_a$ provides a complete description of the conditional distribution. Ideally, we’d have something analogous in the continuous case; i.e. non-parametric, conditional distributions consistent with the conditional expectations from which to draw simulated values. Second, as with any PMM approach, the appropriate value of $k$ is not clear. The literature on preferred $k$ (see Van Buuren\citep{Van2018}, Section 3.4.3) is based on simulation studies and general recommendations. Third, the expense of the $k$NN operation increases with $N_a$, $N_b$, $p$, and $k$. The fusionACS context assumes (at a minimum) large $N_a$, leading to concerns about computation time\footnote{In practice, the fusion operation works with $MN_a$ rows of recipient data, where $M$ is the number of implicates. So, the effective row size passed to the $k$NN operation is $>50$ million for a single year of ACS households given typical $M = 40$.}.

To address these issues, we modify the approach outlined above. First, we find the $K$ nearest neighbors in $D_b$ associated with each observation of $D_b$ (not $D_a$). Since the fusionACS context implies $N_b \ll N_a$, the $k$NN step using $D_b$ is not usually a problem (later we introduce an option for handling even large $N_b$). This yields a $N_b \times K$ matrix (call it $S$) of observed $Z$ values, where each row contains values sourced from donor observations with the most-similar conditional expectations.

Note that the conditional expectations can exhibit widely-varying magnitudes. To ensure that the $k$NN step gives approximately equal weight to each expectation, we scale the columns of the input matrices. If $x$ is column $j$ of input matrix $D$, the transformed values are:
\begin{equation*}
D_j=\frac{\frac{x-med(x)}{mad(x}-\Phi(\epsilon)}{2\Phi(1-\epsilon)}
\end{equation*}
where $med(x)$ and $mad(x)$ are the median and median absolute deviation, respectively, and $\epsilon=0.001$. This results in robust scaled values such that $med(D_j)=0.5$ with range approximately $\{0, 1\}$.

Next, for each row in $S$, we find the unique integer value $k^* (k^* \leq K)$ that yields the best empirical approximation of the conditional distribution of $Z|X_b$. That is, for each row in $S$, we find $k^*$ such that the first $k^*$ values result in mean and quantile values most similar to those in $D_b$. This is done by minimizing an objective function for each row in $S$.

Let $x$ contain the first $k$ values from row $i$ of matrix $S$. We calculate measures of divergence between $x$ and the conditional mean and quantiles ($u$ and $Q_{1:p}$) from row $i$ of $D_b$. The divergence from the conditional mean is:
\begin{equation*}
\Delta_u=1-\frac{\phi(\frac{\bar{x}-u}{\sigma})}{\phi(0)}\text{, where }\sigma=\frac{Q_p-Q_1}{\Phi(P_p)-\Phi(P_1)}
\end{equation*}
We then calculate a measure of divergence for each of $Q_{1:p}$ conditional quantiles:
\begin{equation*}
\Delta_p=\left|\frac{\sum(x \leq Q_p)}{k}-P_p\right|\div\tau
\end{equation*}
where $\tau=P_p$ when $P_p>0.5$ and $\tau=1-P_p$ otherwise.

The overall divergence:
\begin{equation*}
\Delta=\Delta_u+\sum_1^p\Delta_p
\end{equation*}

The deltas are calculated for each value of $k$ in $1:K$, and the optimal $k_i^*$ is that which minimizes $\Delta$. The use of  $\phi$ (normal PDF) and $\Phi$ (normal CDF) do not imply any parametric assumptions about the shape of the conditional distribution itself. The derivation of $\sigma$ from the conditional quantiles assumes a normal distribution\citep{Cook2010}, but this is done only to plausibly scale the mean divergence to $\{0, 1\}$. Note that both of the deltas are bounded $\{0, 1\}$ and equal zero when there is perfect agreement between $x$ and the conditional expectations, allowing them to be summed. Critically, $k_i^*$ can be determined using maximally-efficient matrix operations, even when $S$ is large.

This operation produces a list $(L)$ of $N_b$ variable-length (i.e. variable-$k^*$) vectors of observed $Z$ values that give an empirical approximation of each donor observation’s conditional distribution for $Z|X_b$. For each row of $D_a$, we find the row index $i\in\{1,N+b\}$ of the single nearest neighbor in $D_b$. A simulated value is then randomly drawn from the observed $Z$ values in $L_i$. Finding the single nearest neighbor is fast.

\begin{figure}[ht]
\centering
\begin{subfigure}[b]{0.6\textwidth}
\includegraphics[width=\linewidth]{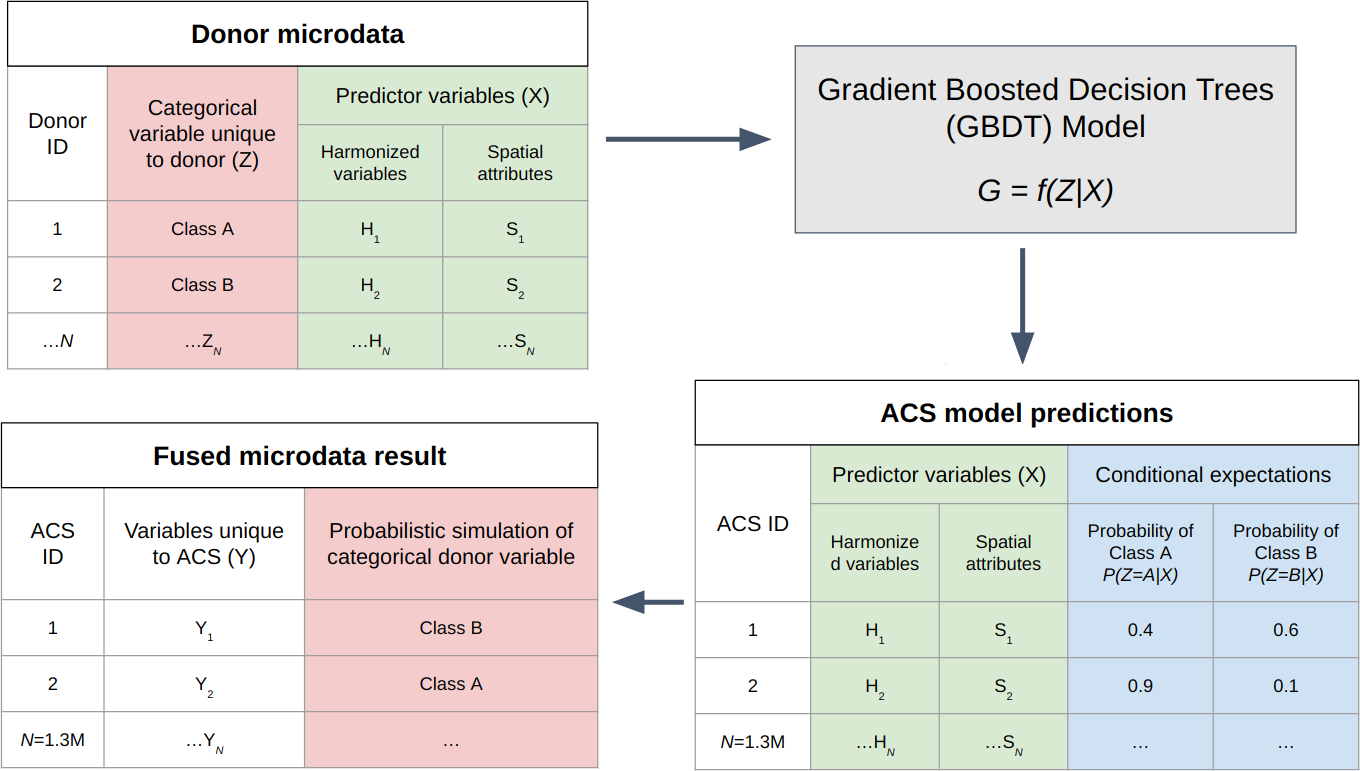}
\caption{Fusion of a single categorical variable}
\end{subfigure}
\begin{subfigure}[b]{0.75\textwidth}
\includegraphics[width=\linewidth]{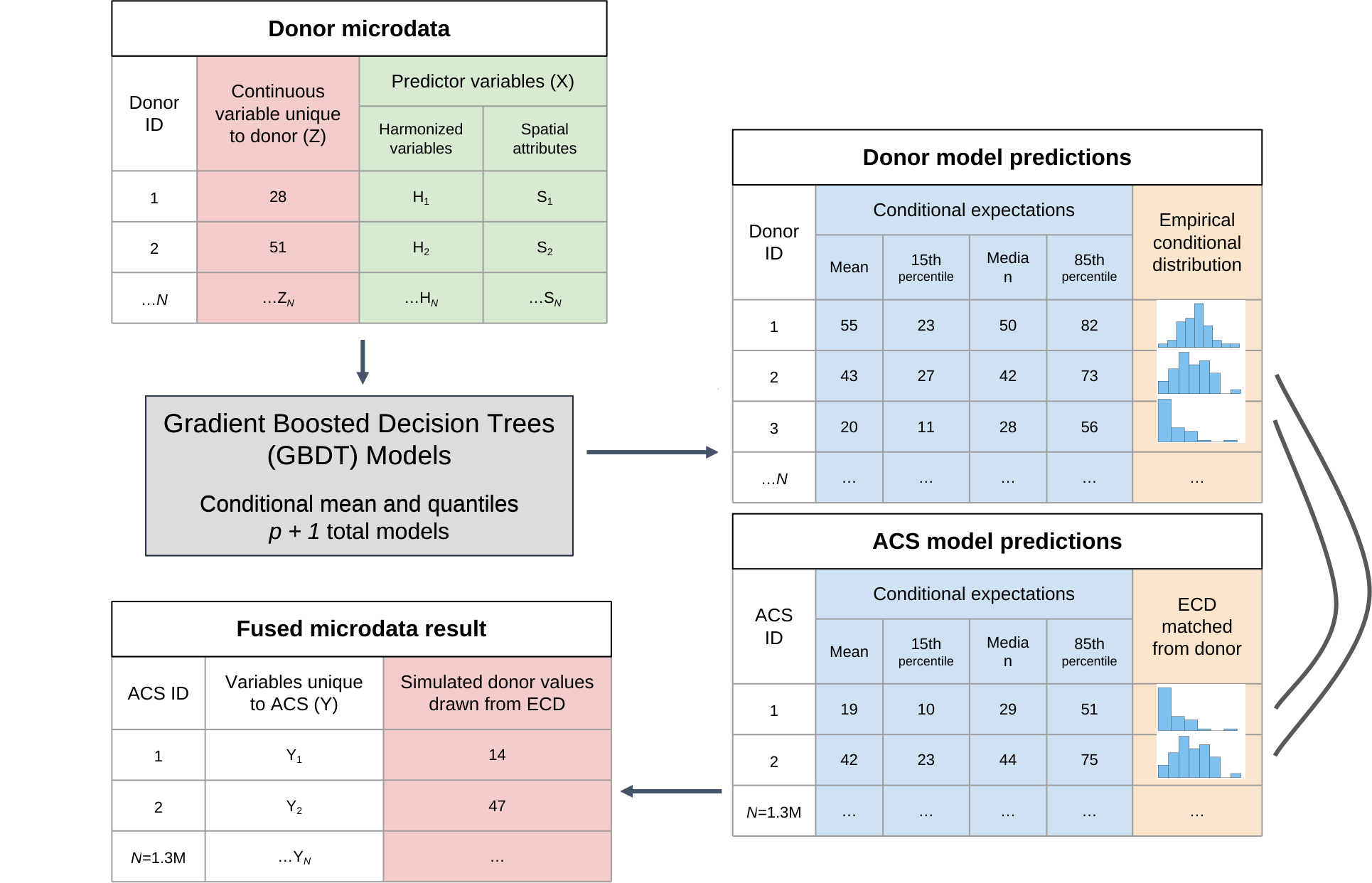}
\caption{Fusion of a single continuous variable}
\end{subfigure}
\begin{subfigure}[b]{0.45\textwidth}
\includegraphics[width=\linewidth]{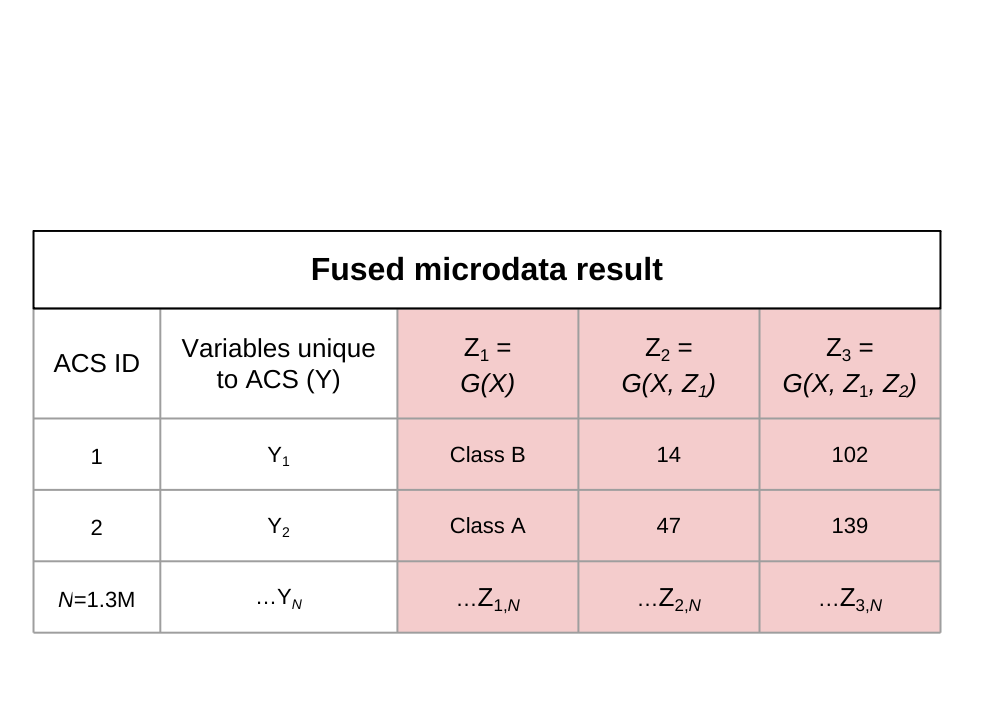}
\caption{Chained fusion}
\end{subfigure}
\begin{subfigure}[b]{0.45\textwidth}
\includegraphics[width=\linewidth]{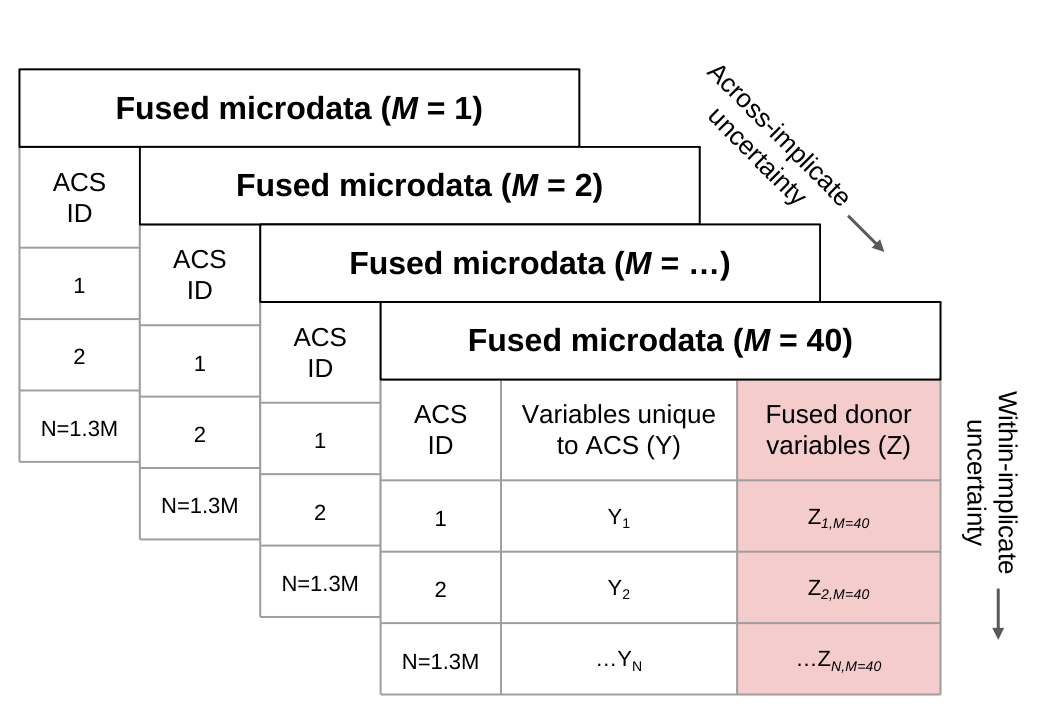}
\caption{Generation of multiple implicates}
\end{subfigure}
\caption{fusionACS: Schematic of the fusion process}
\label{fig:fig_2}
\end{figure}

To recapitulate: For each donor observation, we construct an empirical approximation of the conditional distribution, $Z|X_b$, using observed $Z$ values. Conditional expectations are modeled for each recipient observation. Each recipient is matched to the donor observation with the most similar conditional expectations. Finally, simulated $Z$ values are drawn from the empirical conditional distribution of the matched donor observation. Figure \ref{fig:fig_2} shows schematic diagrams of the process for categorical and continuous variables.

This “variable-$k$” approach has desirable properties: it does not require a fixed $k$; it explicitly uses the conditional expectations to approximate a non-parametric conditional distribution; computation time is not unduly influenced by $N_a$; and the simulated values are drawn from observed $Z$, ensuring valid outcomes.

In principle, it is preferable to use $D_a$ in the initial $K$ nearest neighbors step, resulting in $S$ being a $N_a \times K$ matrix containing observed $Z$ values. However, we find that $k^*$ is typically much larger than the $k = 5$ or $k = 10$ used in conventional PMM. With $k^*$ regularly on the order of 100 to 300, $K$ needs to be large enough to ensure we capture a good approximation of the conditional distribution ($K = 500$ by default in fusionModel). If both $N_a$ and $K$ are comparatively large, the required $k$NN operation may be unduly slow when using $D_a$ directly (as is the case for fusionACS applications).

fusionModel includes an additional option to speed up calculations in the event that $N_b$ is large. In this case, we can first perform $k$-means clustering on $D_b$ to reduce it to some smaller number $(r)$ of cluster centers. With $D_b$ reduced to an $r \times (p + 1)$ matrix, the calculations proceed as above but significantly faster when $r \ll Nb$.

Semi-continuous $Z$ that is inflated at zero is common in social surveys, especially variables related to dollar amounts. We use a two-stage modeling approach in this special case. A categorical (binary) model is first used to simulate zero vs. non-zero outcomes. Then $p + 1$ mean and quantiles models and the variable-$k$ approach described above are used to simulate outcomes, conditional on $Z \neq 0$.

If there are multiple fusion variables, $Z_{1:n}$ , they are fused sequentially such that $G_i = f(Z_i|X,Z_{1:i-1})$. Fusion variables earlier in the fusion sequence become available as predictors. This allows within-observation dependence among the fusion variables to be modeled explicitly (at least for $Z$ that occur later in the sequence), as well as being mediated through $X$.

Sometimes it is useful to fuse variables in “blocks”. This is most relevant when there are fusion variables that are structurally linked. For example, if a set of continuous variables need to sum to one at the household level, they must be fused in a block to ensure this identity is preserved in the output. Variable blocks can contain any variable type (categorical, continuous, semi-continuous). For computational convenience, fusion of blocks employs the fixed-$k$ conditional expectation matching approach first described. That is, $k$ is fixed to some user-specified integer ($k = 10$ by default). In this case, $D_b$ and $D_a$ include the conditional expectations of all variables in the block. If all $Z$ are in a single block, then the fusion process equates to sampling complete records of $Z$ from the donor using fixed $k$.

\subsection*{Modeling details}

Successful fusion hinges on the amount of information that can be extracted from $X$. The Data Preparation section describes how we maximize the amount of potentially useful $data$ in $X$. Our ability to then extract useful $information$ depends critically on the modeling strategy used to estimate $f(Z|X)$.

The fusionACS project uses LightGBM gradient boosting models (GBM)\citep{Ke2017}, because they are flexible and efficient – functionally, computationally, and in terms of predictive ability. By changing the loss function, we can use a single modeling framework for prediction of conditional probabilities, means, and quantiles. GBM’s do not require a specified functional form and make no parametric assumptions. They can handle many predictor variables and automatically detect important predictors, interaction effects, and non-linear relationships. Tuning and cross-validation during training results in models that exhibit state-of-the-art predictive ability. And since LightGBM was designed for large-scale machine learning applications, even comparatively large fusionACS exercises compute efficiently.

Gradient boosting is (largely) a “black box” machine learning strategy ideal for contexts that demand high predictive ability but care little about inference. That is not generally the case in academic settings, but it is a good description of the fusionACS context. Since the platform seeks to accommodate and convincingly model any variable from any donor survey, GBM’s ability to perform well under what we might call “hands off, kitchen sink” conditions is an advantage. 

The primary danger here is that a model could “overfit” to the training data, learning spurious patterns that are a result of random noise instead of legitimate signal. This issue receives little attention in the larger synthetic data literature, because it is largely focused on creating synthetic versions of the donor survey itself; i.e. replication of noise in the donor is not necessarily a problem. In the fusionACS case, the overarching goal is to estimate how ACS respondents might have answered the donor survey questionnaire. This implies learning generalizable patterns in the donor data (i.e. avoidance of overfitting). Or, to put it differently, overfit models will underestimate the amount of variance that we would reasonably expect ACS respondents to exhibit if they actually completed the questionnaire.

To protect against overfitting, we train each LightGBM model using 5-fold cross-validation to find the number of iterations (i.e. number of tree learners) that minimizes the out-of-sample loss metric. The final model is fit to the complete data set using this optimal number of iterations. In addition, we test three different tree sizes (number of leaves: 16, 32, 64), subsample 80\% of predictors in each iteration, and set the minimum number of node observations to 0.1\% of $N_b$ (minimum 20). All of these settings are designed to reduce the risk of overfitting during training. In addition, we employ a “prescreen” step that selects a unique subset of the predictor variables in $X$ to use with each fusion variable in $Z$. This helps reduce both the risk of overfitting and computation time. While there is no penalty to making $X$ as data-rich as possible (in general), we don’t ask the GBM modeling process itself to handle potentially hundreds of predictors. Doing so would unnecessarily increase the chance of a model learning a spurious pattern.

The prescreen step fits LASSO models\citep{Friedman2010} using the complete $X$ and choosing the model that explains 95\% of deviance, relative to a “full” model that includes all potential predictors. Since the LASSO shrinks coefficients towards zero, the selected model utilizes only a subset of $X$, and it is a useful screening strategy in the presence of highly-correlated predictors – as is often the case for fusionACS applications given the large number of correlated spatial attributes present in $X$.

\subsection*{Implementation details}

The methodology described above is implemented in the open-source fusionModel $R$ package as two primary functions: \textit{train()} and \textit{fuse()}. The former encompasses a LightGBM model fitting to a donor dataset and the variable-$k$ calculations, while the latter makes conditional expectation predictions for a recipient dataset and then draws simulated outcomes.

The \textit{train()} function was written to enable maximum speed and memory efficiency via forking on Unix-like systems (e.g. Linux servers). On Windows machines, OpenMP-enabled multithreading is used within the LightGBM model training step only (forking is not possible on Windows). We have found forking to be faster for typical donor microdata, and this is what we use for production runs on Linux servers.

The \textit{fuse()} function takes advantage of LightGBM’s native multithreading regardless of platform, since the expensive step is prediction of the numerous GBM’s for the ACS recipient microdata. To accommodate intended fusionACS applications, \textit{fuse()} intelligently “chunks” operations depending on available system memory and writes output to disk “on the fly”. This makes large-scale fusion tasks possible (even if they cannot fit in physical RAM) and allows the multithreading to operate near peak efficiency.

Both \textit{train()} and \textit{fuse()} include an approximate nearest neighbor search, for which they use the ANN library\citep{Mount2010} implemented via the RANN package\citep{Jefferis2020}. LASSO models are fit using the glmnet package\citep{Friedman2010,Simon2011}. More generally, fusionModel relies on the data.table\citep{Dowle2022} and matrixStats\citep{Bengtsson2017} packages for the key data manipulation steps. All of these packages – as well as LightGBM – are maximized for efficiency and written in low-level C code. So even though fusionModel itself is written in $R$, the vast bulk of the computation is optimized for speed and memory usage.

\subsection*{Uncertainty estimation}

The fusion process attempts to produce a realistic representation of how each ACS respondent household (or individual) might have answered the questionnaire of the donor survey. The fused values are inherently probabilistic, reflecting uncertainty in the underlying statistical models.

In order to fully capture this uncertainty, fusionACS output consists of $M$ multiple "implicates". A single implicate contains a simulated response for each fused variable and ACS-PUMS respondent. Each implicate provides a unique, plausible set of simulated outcomes. Multiple implicates are needed to calculate unbiased point estimates and associated uncertainty (margin of error) for any particular analysis of the data, making it the standard approach in the literature\citep{Rubin1996}.

The use of multiple implicates is conceptually akin to that of replicate weights in conventional survey analysis. Replicate weights quantify uncertainty (variance) by keeping the response values fixed but varying the weight (frequency) associated with each respondent. Conversely, when imputing (or fusing) data, the primary sample weights are typically fixed while the simulated values vary across implicates.

Since proper analysis of multiple implicates can be rather cumbersome – both from a coding and mathematical standpoint – the fusionModel package provides a convenient \textit{analyze()} function to perform common analyses on fused data and report point estimates and associated uncertainty. Potential analyses currently include variable means, proportions, sums, counts, and medians, (optionally) calculated for population subgroups.

Point estimates for any particular analysis are simply the mean of the $M$ individual estimates across the implicates. In general, higher $M$ is preferable but requires more computation and larger output file size. For fusionACS production runs, we currently use $M = 40$ as a reasonable compromise.

Uncertainty for a given estimate reflects standard errors “pooled” across the implicates. A number of pooling rules for implicates have been introduced in the imputation and synthesis literatures, beginning with that of Rubin\citep{Rubin1987} for multiple imputation contexts. The closest analog to the fusionACS context is that considered in Reiter\citep{Reiter2008}. Unfortunately, the pooling formulae in Reiter\citep{Reiter2008} assume a two-stage simulation strategy with parametric models that is not straightforward to apply to fusionACS output. However, that paper shows that the original Rubin\citep{Rubin1987} pooling formulae result in somewhat positively biased variance compared to the “correct” formulae. Consequently, the fusionModel \textit{analyze()} function uses the Rubin\citep{Rubin1987} method to conservatively estimate uncertainty and associated margin of error (MOE). The MOE returned by \textit{analyze()} reflects a 90\% confidence level, consistent with how the Census Bureau reports MOE for native ACS-based estimates.

The $unpooled$ standard errors (SE’s) that are used within the pooling formulae are calculated using the variance within each implicate. For means (and sums), the ratio variance approximation of Cochran\citep{Cochran1977} is used, as this is known to be a good approximation of bootstrapped SE’s for weighted means\citep{Gatz1995}. For proportions, a generalization of the unweighted SE formula is used. For medians, a large-$N$ approximation is used when appropriate\citep{Doob1935} and bootstrapped SE’s computed otherwise.

The \textit{analyze()} function can also (optionally) include uncertainty due to variance in the ACS PUMS replicate weights. This is generally preferable, since it captures uncertainty in both the fused (simulated) values and the sampling weight of ACS households within the population. We find that including replicate weight uncertainty often increases MOE’s by 15-30\%. We introduce replicate weight uncertainty by assigning a different set of replicate weights to each of the $M$ implicates (there are 80 PUMS replicate weights, so we use half of the replicate weights when $M = 40$). We then estimate the additional across-implicate variance when using replicate weights (compared to the primary weights), and add this to Rubin's pooled variance.

\section*{Data Records}

The fusionACS outputs generated include:

\begin{itemize}[noitemsep]
\item Fusion of 12 select donor variables from RECS 2015 to ACS 2015. 
\item Fusion of 5 select donor variables from AHS 2019 to ACS 2019.
\item Fusion of 5 select donor variables from NHTS 2017 to ACS 2017.
\item Fusion of 47 household consumption-expenditure and tax variables from CEI 2015-2019 (pooled) to ACS 2019.
\end{itemize}

The fused RECS 2015, AHS 2019, and NHTS 2017 microdata  consists of a a single .fst file per survey, each containing 40 implicates. Important to notice, the RECS expenditure variables were fused in a second step using only the consumption  and location variables as predictors to attain local consistency in energy prices. 

The fused CEI 2015-2019 microdata as well as PUMA-level estimates for select variables consists of a .fst file containing 30 implicates. The data include a single “tax” variable derived from the CEI’s native before- and after-tax income variables.

\begin{figure}[ht]
\centering
\begin{subfigure}[b]{0.8\textwidth}
\includegraphics[width=\linewidth]
{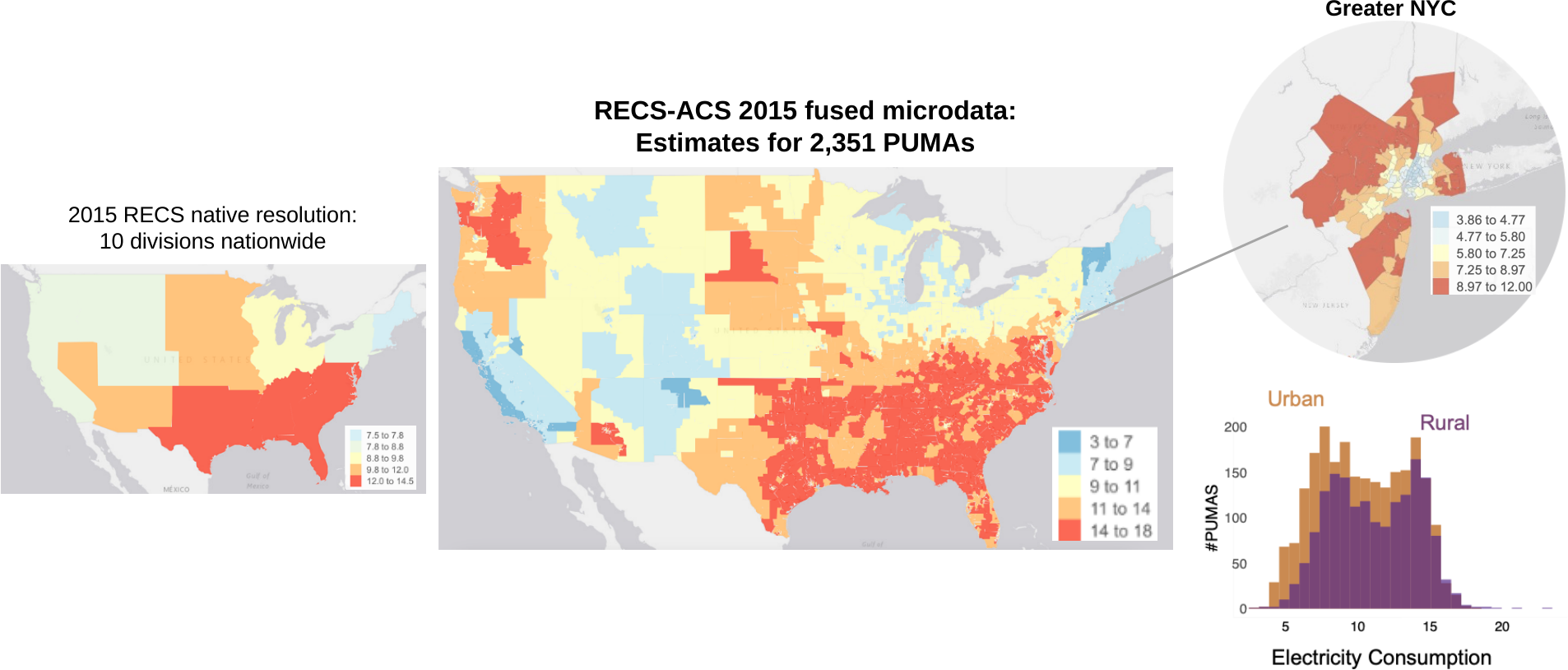}
\caption{Annual Electricity Consumption (MWh) in RECS vs RECS-ACS fusion}
\label{fig:fig_3_1}
\end{subfigure}
\begin{subfigure}[b]{0.8\textwidth}
\includegraphics[width=\linewidth]
{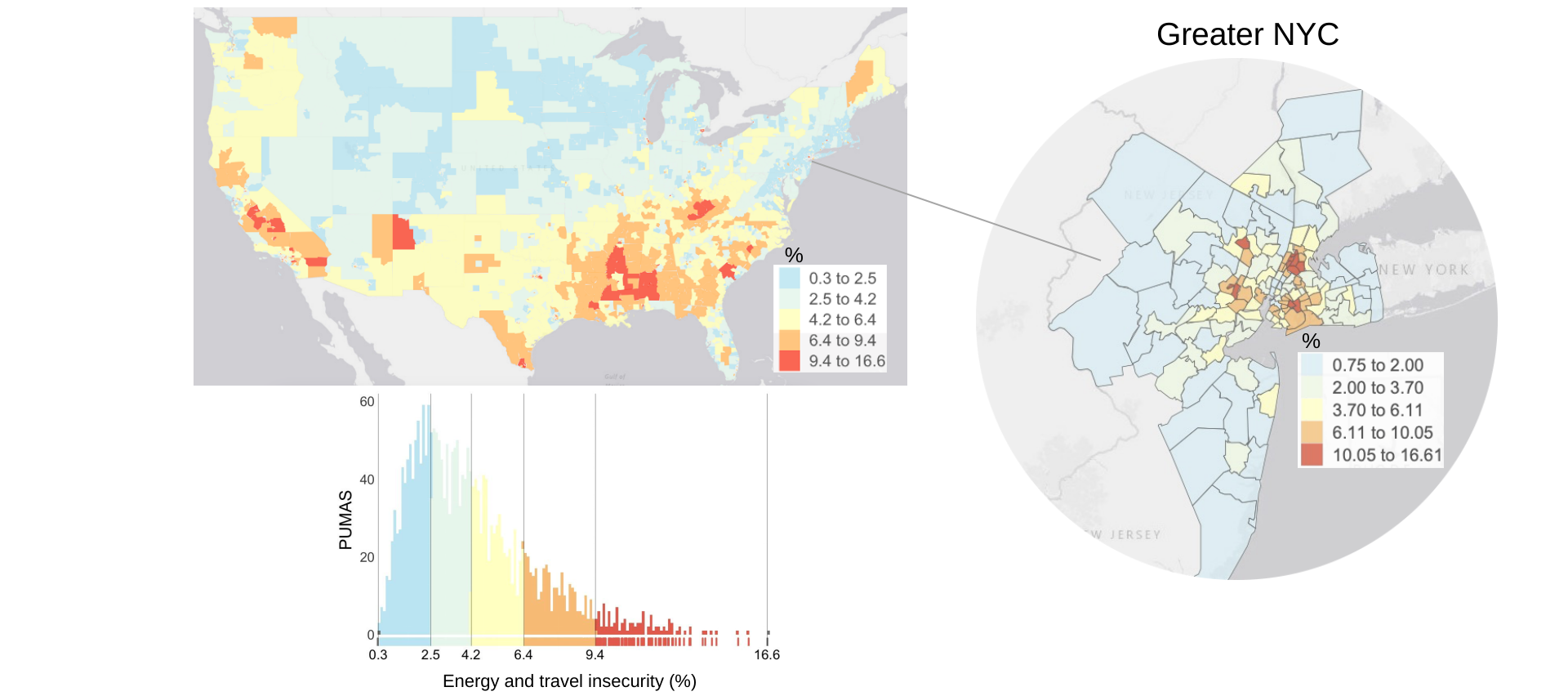}
\caption{Households experiencing both energy (RECS-ACS) and travel insecurity (NHTS-ACS)}
\label{fig:fig_3_2}
\end{subfigure}
\caption{fusionACS: examples of enhanced capabilities of the fused microdata}
\label{fig:fig_3}
\end{figure}

The fused variables were selected to test functionality across variable types and provide examples of socially-relevant  variables. The list of fused variables and their description can be found in Table \ref{tab:tab_2}. Graphical examples of some of the capabilities of fusion outputs can be seen in Figure \ref{fig:fig_3}, including higher spatial granularity (Figure \ref{fig:fig_3_1}) and the potential to do multidimensional analyses using variables from multiple surveys (Figure \ref{fig:fig_3_2}).

\subsection*{Data Preparation}
A significant amount of effort is required to prepare raw survey microdata so they can be used within the fusionACS pipeline. The production of standardized microdata, harmonized variables, spatial datasets, and associated documentation across multiple surveys is a major contribution of the fusionACS project. The code used to achieve this is housed in the fusionData github repository\citep{Ummel2022_1}. 

\subsubsection*{Ingestion}

“Ingestion” of a survey requires transforming raw survey data into standardized microdata that meet certain requirements. This entails writing custom code for every donor survey, vintage, and respondent type (household and/or person), often involving tedious “pre-processing” tasks like (among others things): replacement of integer codes with descriptive variable levels; replacement of “valid blanks” and “skips” with plausible values; imputation of missing observations; appropriate “classing” of variables (e.g. defining ordered factors); and documentation of variables. This process sometimes identifies errors or irregularities in the raw survey data, which suggests that we are thoroughly interrogating the data during the ingestion step.

Ingestion results in processed microdata observations that meet the following conditions:
\vspace{-0.1cm}
\begin{itemize}[noitemsep]
\item Contains as many observations and variables as possible.
\item Variable names and descriptions are taken from the official codebook, possibly modified for clarity.
\item Official variable names are coerced to lower-case alphanumeric, possibly using single underscores.
\item Codes used in the raw data are replaced with descriptive labels from the codebook; e.g. integer values are replaced with associated factor levels.
\item All “valid blanks” in the raw data are set to plausible values; NA’s are often actual zeros or some other knowable value based on the question structure.
\item All “invalid blanks” or missing values in the raw data are imputed.
\item Ordered factors are used and defined whenever possible (as opposed to unordered).
\item Standard column names are used for unique household identifiers (e.g. “acs\_2019\_hid”); for person-level microdata the within-household person identifier (integer) is always “pid”.
\item Standard column names are used for observation weights; “weight” for the primary weighting variable and “rep\_1”, etc. for replicate weights.
\end{itemize}

\subsubsection*{Harmonization}

Once a donor survey has been successfully ingested, it can then be “harmonized” with the ACS in preparation for fusion. The harmonization step identifies variables common to both the donor survey and the ACS and is the statistical linchpin of the fusion process. It consists of matching conceptually similar variables across surveys and determining how they can be modified to measure similar concepts. The harmonization process should be as exhaustive as possible, since the predictive power of subsequent LightGBM models depends on the amount of information in the shared/harmonized predictor variables.

\begin{figure}[ht]
\centering
\fbox{\includegraphics[width=0.95\linewidth]{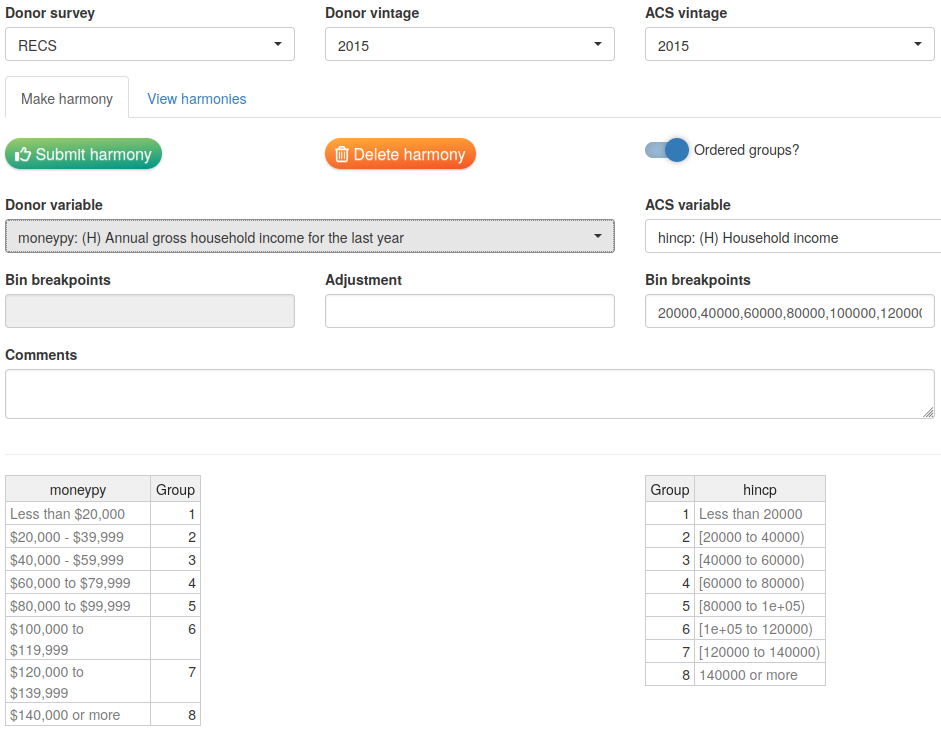}}
\caption{fusionACS: Example of the harmonization tool depicting the harmonization between the categorical variables \textit{moneypy} from RECS and the continuous variable \textit{hincp} from ACS.}
\label{fig:fig_4}
\end{figure}

In general, harmonization is time-consuming and error-prone. To address these concerns, we built a custom “Survey Harmonization Tool” – a web app within the fusionData package – to make the harmonization process faster and more robust. The harmonization tool can define complex harmonies, including “one-to-one”, “many-to-one”, and “many-to-many” linkages, as well as “binning” (discretization) of continuous variables in one survey to create alignment with a categorical variable in another. Figure \ref{fig:fig_4} is a screenshot of the harmonization tool in which the ACS continuous household income variable (“hincp”) is binned to create harmony with the RECS categorical “moneypy” variable.

Importantly, the harmonization process makes use of both household- and person-level variables, when available. This is true even if fusion occurs only at the household level. For example, it is common for donor surveys that solicit person-level information to ask for the age of each household member. This variable can be harmonized with an analogous variable in the ACS person-level microdata. Even if the eventual fusion step models and simulates household-level variables for each ACS respondent household (as is typical), the person-level harmonies are still utilized. In this case, the underlying code automatically constructs a household-level variable reporting the age of the householder/reference person (constructed from the person-level microdata and associated harmonies). In this way, we leverage maximum information that is common to both the donor and ACS.
There is often considerable conceptual overlap between donor survey variables and those in the ACS. For example, for the RECS, we have identified 24 harmonies such as matching the household race variable that consists of 6 categories in the RECS with the 9 categories in the ACS, creating 8 breakpoints in the household income variable from the ACS to match with the corresponding household income variable in the RECS which consists of 8 categories. These variables are key to the explanatory power of the models used in the fusion process.

\subsubsection*{Spatial variables}

Part of the data processing involves adding spatial variables to the donor and ACS microdata to expand the number of predictor variables available for the modeling step. Spatial variables help to characterize a household’s location/environment, as opposed to the respondent-specific characteristics used in the harmonization process. For example, knowing something about the population density of a household’s general location can help explain patterns in the variables being fused that might not be “picked up” by a model using only respondent characteristics.

Spatial predictor variables come in two flavors: “disclosed” and “third-party”. Disclosed spatial variables are typically location identifiers that the donor survey administrators have deemed safe to disclose (e.g. metropolitan area, state, region, climate zone). We can use these variables “as-is”, since they only require that the same variable can be constructed for ACS respondents (usually straightforward).

In principle, there is no limit to the amount, nature, or resolution of third-party spatial information that can be utilized by the fusionACS platform. The only requirement is that a spatial dataset must have national coverage. To date, we have focused on readily-available datasets likely to be useful in explaining the kind of socioeconomic phenomena measured by the donor surveys ingested so far (see Table \ref{tab:tab_1}). The number and range of potential spatial datasets is much greater than this; ingesting additional datasets simply hasn’t been a priority up to this point.

Spatial variables are merged to the donor and ACS microdata at the level of individual PUMA’s. This is because the ACS PUMS only discloses respondent location for PUMA’s, so this is as precise as we can be with the spatial variables. For example, the EPA-SLD dataset provides variables describing features of the built environment for individual block groups. These variables are summarized at the PUMA-level prior to merging to the donor and ACS microdata. They are then available as LightGBM predictor variables in both the training (on donor microdata) and prediction (on ACS microdata) steps.

Due to confidentiality constraints, all of the donor surveys ingested so far do not disclose the PUMA of respondents. Consequently, we impute each donor respondent’s PUMA given observable information. We make use of disclosed location information as well as the suite of harmonized respondent-level variables. The latter are used to perform a probabilistic imputation, using Gower’s distance\citep{Gower1971} as a similarity measure between donor and ACS respondents. That is, we assign a PUMA to each donor respondent by matching to an ACS respondent (using its observed PUMA) within the same disclosed location (e.g. state), where the probability of selection is proportional to the similarity of the donor and respondent on observable (harmonized) characteristics. This allows us to leverage all available information in the imputation process.

Figure \ref{fig:fig_3_1} presents an example of the higher spatial granularity that can be achieved as an outcome of the fusion methodology, starting from the 10 large US Census divisions available in RECS, to the 2,351 PUMAs available in ACS.

\section*{Technical Validation}

Prospective users of fusionACS require information about the quality or validity of the data outputs. In practice, analysts want to know if a fused dataset or variable is “good enough”. There are a number of ways to answer this question, depending on what is meant by “good enough” and the nature of the desired analysis.

The most general answer is that any analysis of fused data can and should make use of the multiple implicates to calculate uncertainty (margin of error) alongside point estimates of interest. The margin of error should be taken into consideration when deciding if the results of an analysis are “good enough” for the intended application. This is best practice for any analysis of survey microdata, fused or otherwise.

The \textit{analyze()} function in the fusionModel package was built for this purpose and enables users to correctly calculate means, proportions, sums, counts, and medians, (optionally) for population subsets (e.g. by race). See the “Uncertainty estimation” section for details. There are always limits to how much one can ask of the available data, and uncertainty estimation is the principal tool for detecting when an analysis has “gone too far” – the definition of which can only be specified by the analyst.

That said, prospective users may wish to know “how good” the fusion outputs are prior to performing any actual analysis; i.e. How do we know that the resulting point estimates and standard errors are plausible? What kind of accuracy can one reasonably expect for different kinds of analyses?

These questions necessitate validation exercises to demonstrate the expected accuracy or “utility” of fusion-based analyses. In the data synthesis literature, a distinction is drawn between “general” (global) and “specific” (narrow) utility of synthetic datasets\citep{Snoke2018}. The former provides an overall statistical measure of the similarity of synthetic and observed data, while the latter refers to specific comparisons of synthetic and observed data for the kinds of analyses that users actually deploy in practice (means, coefficients, etc.). General utility is a rather low bar to clear and (more importantly) it does not provide the kind of intuitive and familiar “proof” of data quality that can inspire confidence in prospective users. Consequently, we focus on validation exercises using measures of specific utility.

We also need to draw a distinction between “internal” and “external” validation. Internal validation consists of analyzing synthetic data produced by fusing variables “back” onto the original donor microdata. It is analogous to assessing model skill by comparing predictions to the observed training data. External validation, on the other hand, consists of comparing specific results produced using fused data with analogous results from an entirely independent data source. External validation is difficult in the case of fusionACS, where the greatest “value added” comes from high-resolution or otherwise unique analyses for which no validation data exist.

\subsubsection*{Internal validation}

The fusionModel package includes a \textit{validate()} function to perform specific (non-general) internal validation tests on synthetic variables that have been fused back onto the original donor data. The objective of the \textit{validate()} function is twofold: 1) to confirm that the fusion and analysis code is working as expected (i.e. “sanity check”) and 2) to convey some sense of the expected utility of the synthetic data across potential analyses.

Utility in this case is based on comparison of analytical results derived using the multiple-implicate fusion output with those derived using the original donor microdata. By performing analyses on population subsets of varying size, \textit{validate()} estimates how the synthetic variables perform for analyses of varying difficulty/complexity. It computes fusion variable means and proportions for subsets of the full sample – separately for both the observed and fused data – and then compares the results. The user specifies the non-fusion variables used to construct subsets. For example, \textit{validate()} might be called using education and race to construct population subsets (i.e. by education, by race, and by education-race) for which estimates of the fusion variables are calculated and then compared.

Smaller subsets are noisier and more susceptible to outliers in the observed data. Consequently, the discrepancy between observed and simulated estimates tends to increase in magnitude and become noisier as subset size declines. For validation purposes we want to know what the general trend looks like, ignoring noise/outliers in the observed data. The \textit{validate()} function plots results using a median smoother, in order to convey the expected typical (median) performance at a given subset size. It reports results for three different performance metrics, explained below.

\begin{figure}[ht]
\centering
\begin{subfigure}[b]{0.45\textwidth}
\includegraphics[width=\linewidth]{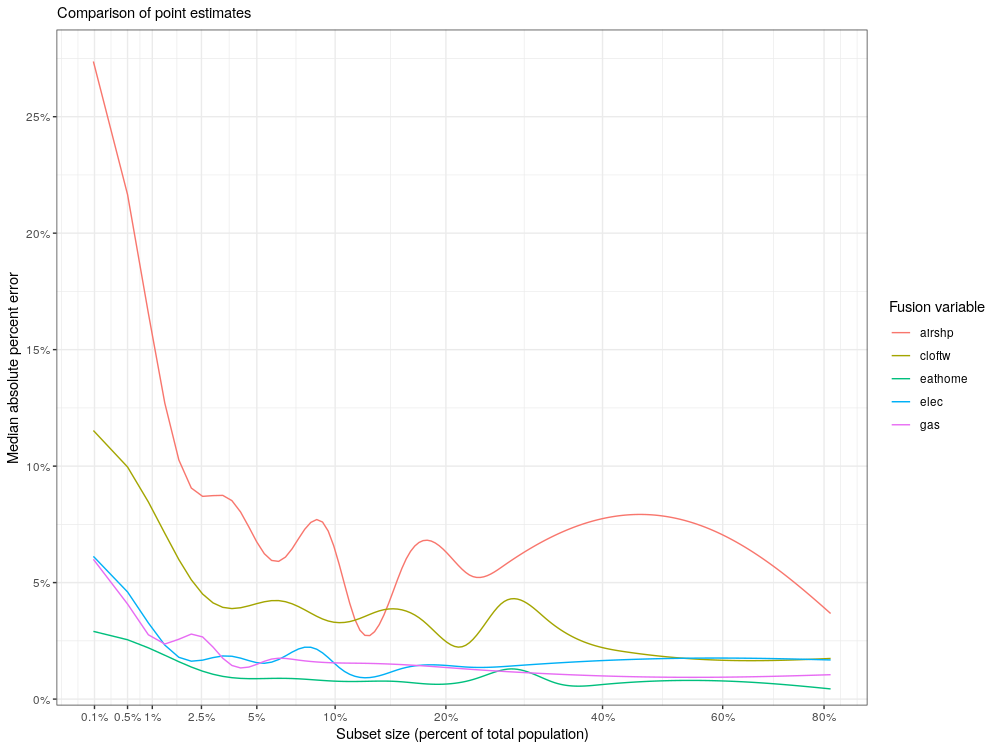}
\caption{Median absolute percent error of point estimates. Variables “eathome”, “elec”, and “gas” exhibit relatively low error, implying that they are well-captured by the underlying models. Variables “cloftw” and “airshp” exhibit higher error, suggesting caution may be warranted when analyzing in detailed analyses.}
\label{fig:fig_5_1}
\end{subfigure}
\quad
\begin{subfigure}[b]{0.45\textwidth}
\includegraphics[width=\linewidth]{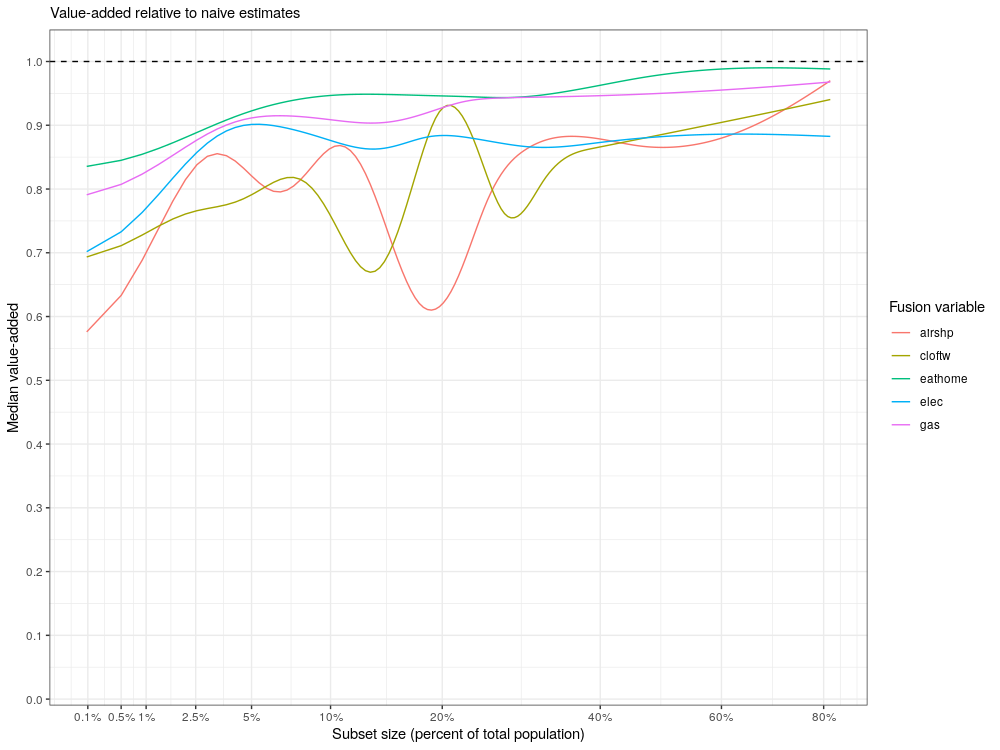}
\caption{Median "value-added" based on comparison of fusion point estimates to those of a naive (null) model. Value-added is generally strong (> 0.8) across all five variables, though it is noisier for “airshp” and “cloftw”.}
\label{fig:fig_5_2}
\end{subfigure}

\begin{subfigure}[b]{0.45\textwidth}
\includegraphics[width=\linewidth]{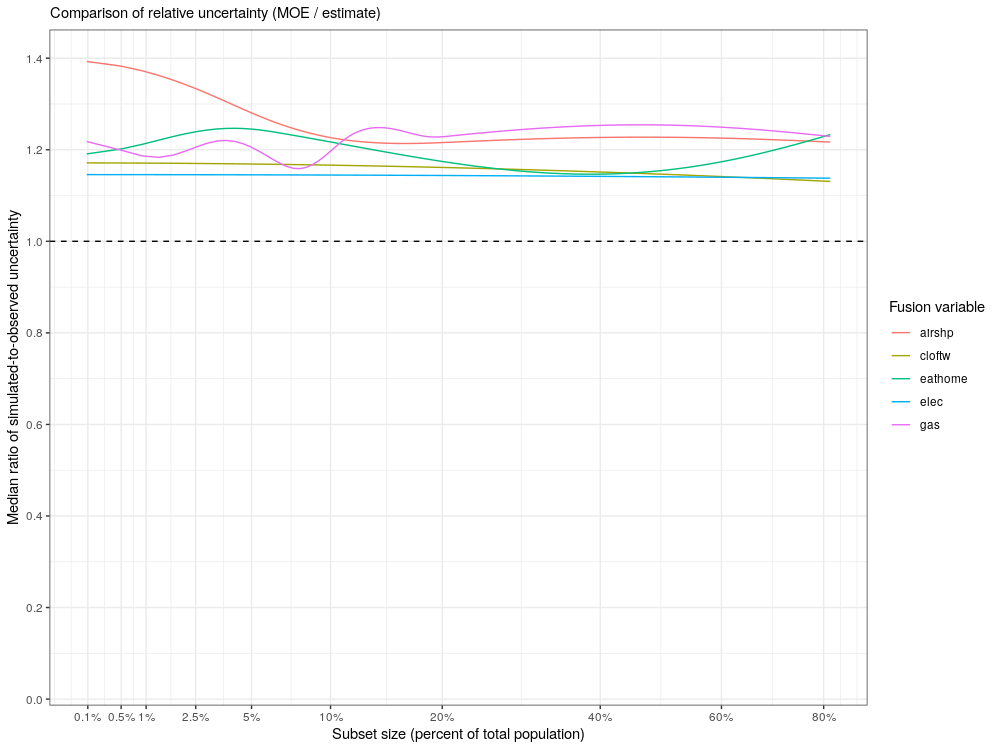}
\caption{Median ratio of fusion point estimate uncertainty to that of original donor. The fusion point estimates typically result in about 20\% higher uncertainty, reflecting the additional uncertainty associated with the modeling process.}
\label{fig:fig_5_3}
\end{subfigure}
\caption{Example internal validation plots for five household expenditure variables from CEI 2015-2019. The results illustrate how fused variables compare to the original donor variables across population subsets of varying size. The variables are: "airshp" (Air and ship travel), "cloftw" (Clothing and footwear), "eathome" (Food eaten at home), "elec" (Electricity), and "gas" (Gasoline).}
\label{fig:fig_5}
\end{figure}

For fusionACS production runs, the fusion models are simulated back onto the donor data and the result passed to \textit{validate()}. The illustrative plots in Figure \ref{fig:fig_5} show \textit{validate()} output for five household expenditure variables fused from the CEI 2015-2019 donor survey. The population subsets are constructed using the six predictor variables that provide the closest analogs for income; race/ethnicity; education; household size; housing tenure; and respondent location. 
Figure \ref{fig:fig_5_1} shows how the observed and simulated point estimates compare, using median absolute percent error as the performance metric. We consider this the easiest-to-interpret error metric for practical purposes. Note that the x-axis is not linear; it is scaled to reveal more detail for small population subsets. The y-axis gives the (smoothed) median absolute percent error at each subset size.

The discrepancy (error) between the observed and simulated point estimates exhibits the typical pattern, increasing as subset size declines, but there is considerable variation. The variables “eathome”, “elec”, and “gas” exhibit quite low percent error, even for small subsets, implying that the explanatory patterns driving these variables are strongly identified by the underlying, cross-validated LightGBM models. The variables “cloftw” (clothing and footwear) and “airshp” (air and ship travel) exhibit higher error, especially for smaller subsets. These results suggest caution might be warranted if using “airshp” (and possibly “cloftw”) in high-resolution or complex analyses.

The \textit{validate()} output plot in Figure \ref{fig:fig_5_2} presents an alternative way to gauge fusion quality, using a “value-added” metric that compares fusion output to that of a naive (null) model. Given simulated point estimate $y_s$ and observed estimate $y_o$, we define the value-added ($V$) as:

\begin{equation*}
V=max\left(0, 1 - \frac{|y_s - y_o|}{|E(y_o) - y_o|}\right)
\end{equation*}

where $E(y_o)$ is the full-sample mean. That is, $V$ measures the extent to which the simulated estimates out-perform the naive estimate of a null model. This is conceptually similar to the approach used to define the canonical coefficient of determination ($R^2$)\footnote{Note that $V$ uses absolute instead of squared error. Consequently, $V$ is lower than $R^2$ , all else equal.}. $V=1$ when $y_s=y_o$ and $V=0$ when $y_s$ is worse than the naive estimate. $V$ is calculated for each individual analysis and then the median smoother applied.

In this case, we observe generally high value-added ($>0.8$) throughout most of the subset size range, though it is noisier for “airshp” and “cloftw”. Value-added helps to isolate the performance of the underlying fusion process, controlling for the degree of variance across population subsets. In some cases, percent error (Figure \ref{fig:fig_5_1}) may be relatively high, but the value-added is also quite high. An analyst may still decide to use the fused variable, on the grounds that the underlying modeling process is performing close to optimal; i.e. this is probably “as good as it gets” for the variable in question, given the available survey data.

Finally, \textit{validate()} outputs a comparison of simulated and observed $relative$ uncertainty (MOE divided by the point estimate). This is useful for confirming that the simulated margin of errors exhibit plausible magnitudes. Figure \ref{fig:fig_5_3} indicates that the fused data typically result in MOE about 20\% higher than we observe in the training data, reflecting the additional uncertainty associated with the modeling process. The “airshp” MOE’s inflate for smaller subsets, reflecting the relative difficulty (also observed in the other plots) in modeling air and ship travel expenditures given the available predictor variables.

\subsubsection*{External validation}

We further validated the RECS fusion output with household electricity and natural gas consumption data at the county level available both publicly and from utility companies for the states of New York (NYSERDA)\citep{NYSERDA}, California (California Energy Commission)\citep{CalEnergyCom}, and Massachusetts (masssavedata)\citep{masssavedata} (Figures \ref{fig:fig_6_1} and \ref{fig:fig_6_2}). In general, we observe a good correlation between the simulated metrics and metrics obtained externally for the total electricity and natural gas consumption metrics and that the ranked metrics for the 3 states are largely preserved. Also, the averaged metrics for electricity and natural gas consumption showed larger deviation (Figures \ref{fig:fig_6_3} and \ref{fig:fig_6_4}), especially for the counties in New York, due to the lower sampling of RECS in these areas. 

We also compared publicly available 311 distress call data made during 2014-2016 reporting heat and/or hot water complaints in New York City (Heat/Hot Water) obtained from NYC Open Data\citep{NYC_open_data}, Austin (Building A/C \& Heating Issues) City of Austin\citep{Austin_311}, Philadelphia(Heat) obtained from OpenDataPhilly\citep{OpenDataPhilly}, and Boston (Heat - Excessive  Insufficient or Heat/Fuel Assistance) obtained from Analyze Boston\citep{AnalyzeBoston} to the energy insecurity burden metric \textit{insec} obtained from RECS-ACS fusion for 2015 at the PUMA-level (described in Table \ref{tab:tab_2}) (Figure \ref{fig:fig_6_5}). Here we find reasonable agreement (corr.coeff = 0.31-1.00) between the rank of insecurity and the number of 311 distress calls which further validates the strength of the fused energy insecurity indicator. 

\subsection*{Applications}
These synthetic data offer a new spatially granular characterization of American households' multidimensional well-being and their living conditions. These data have the potential to advance research on multidimensional poverty and improve justice-oriented policy design.

Existing public surveys offer disconnected, specialized knowledge about US households at a coarse spatial scale. Building a picture, beyond income poverty, of people's living standards, well-being and health, would require piecing together information from disparate sources, but without enough granularity to build a comprehensive picture for any community. The ACS offers limited information in this direction. FusionACS provides the capability to build such a comprehensive picture, with only the available data sources being the limit (e.g. Figure \ref{fig:fig_3_2}). Further, the high spatial granularity of homes and household inhabitants can be overlaid with spatial maps of environmental hazards and infrastructure, enabling a picture also of how provisioning systems and climate/geographic conditions may play into peoples' choices, lifestyles and constraints. This can advance our understanding of the nature and drivers of multi-dimensional poverty.

The Biden Administration's Justice40 Initiative and Executive Orders 13985 and 14008 aim to advance racial justice and support for underserved communities, in part by directing the benefits of public investments to 'disadvantaged communities' (DAC). Current DAC definitions have two limitations: beyond income poverty and health outcomes they lack other indicators of social disadvantage, such as access to essential services. FusionACS offers all the variables that are available in multiple specialized donor surveys, which include housing conditions, utilities, transport, expenditure burdens such as food and health, and many others. Second, fusionACS provides a basis to estimate compound vulnerabilities at a household level. Today, governments define DACs as census tracts with high population shares that suffer multiple social and environmental vulnerabilities, but that are independently derived from disconnected surveys. As such, they assume homogeneity in the experience of disadvantage within these census tracts. FusionACS provides potentially more accurate characterizations, albeit with some statistical uncertainty, of populations that, for instance, claim to suffer "heat or eat" trade-offs, or that live in 'food deserts' and have high mobility burdens, measured in money and time, and numerous other combinations of sociodemographic characteristics and deprivation measures. This would be particularly valuable when the threshold population share required to classify as a DAC is not the overwhelming majority. Further, with a broader range of constituent indicators, fusionACS may be applied to a broad range of targeted public investments in particular domains/sectors that require attention to particular combinations of deprivations. 

\section*{Code and data availability}

The data preparation codes and the specific codes to generate the fused datasets presented in this study are on the fusionData github repository\citep{Ummel2022_1}. The generalized codes for the fusion, analysis, and validation of the datasets are available on the fusionModel github repository\citep{Ummel2022_2}. 

All generated fusion outputs are available as .fst files containing 40 implicates along with the household id from the corresponding ACS, which can be accessed using the fst $R$ package\citep{Klik2022}. Instructions on how to use the various functions of the fusionModel package are available in the corresponding github repository\citep{Ummel2022_2}.

\bibliographystyle{plainnat}
\bibliography{sample}

\section*{Acknowledgements} 

This research was supported by funding from the Environmental Protection Agency through RTI International with Grant GR114933 RTI / EPA and the National Science Foundation Growing Convergence Research Grant GR117886.

\section*{Author contributions statement}

K.U., N.G., D.A.C. and N.D.R. conceived the initial framework. K.U., M.P.-C., K.A., N.G., H.A., C.K. and S.H.T. collected and prepared data. K.U. implemented the model, with M.P.-C. and N.G. providing methodological feedback. K.U., M.P.-C. and N.D.R. wrote the manuscript, with figures provided by K.U. and K.A. All authors reviewed the manuscript.

\section*{Figures \& Tables}

\begin{figure}[ht]
\centering
\begin{subfigure}[b]{0.7\textwidth}
\includegraphics[width=\linewidth]{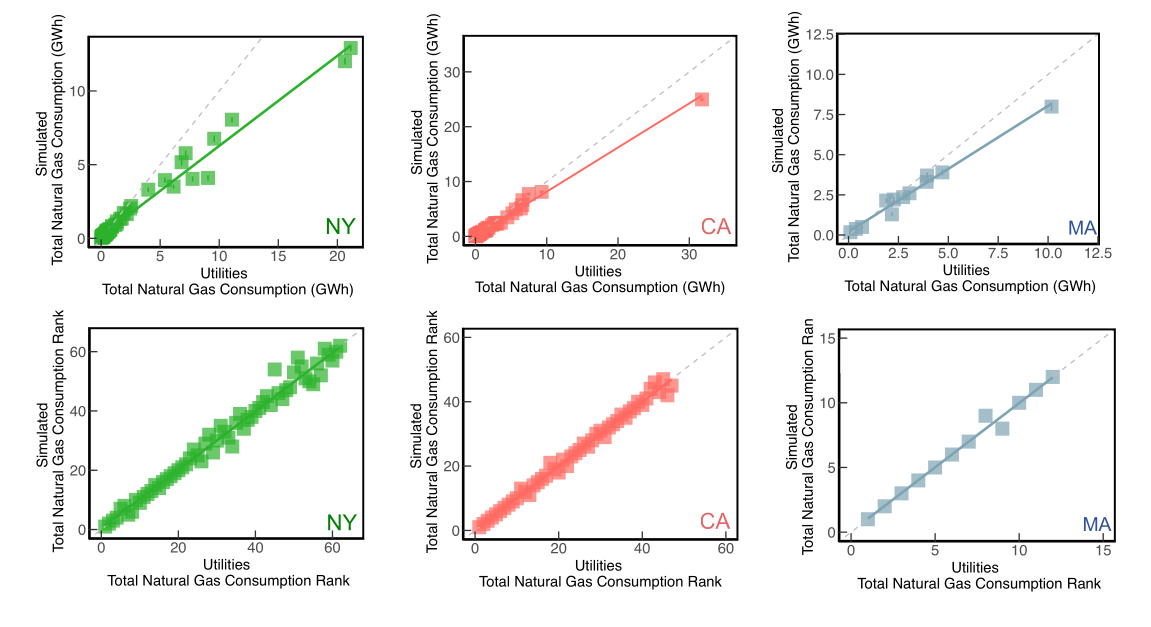}
\caption{}
\label{fig:fig_6_1}
\end{subfigure}
\begin{subfigure}[b]{0.7\textwidth}
\includegraphics[width=\linewidth]{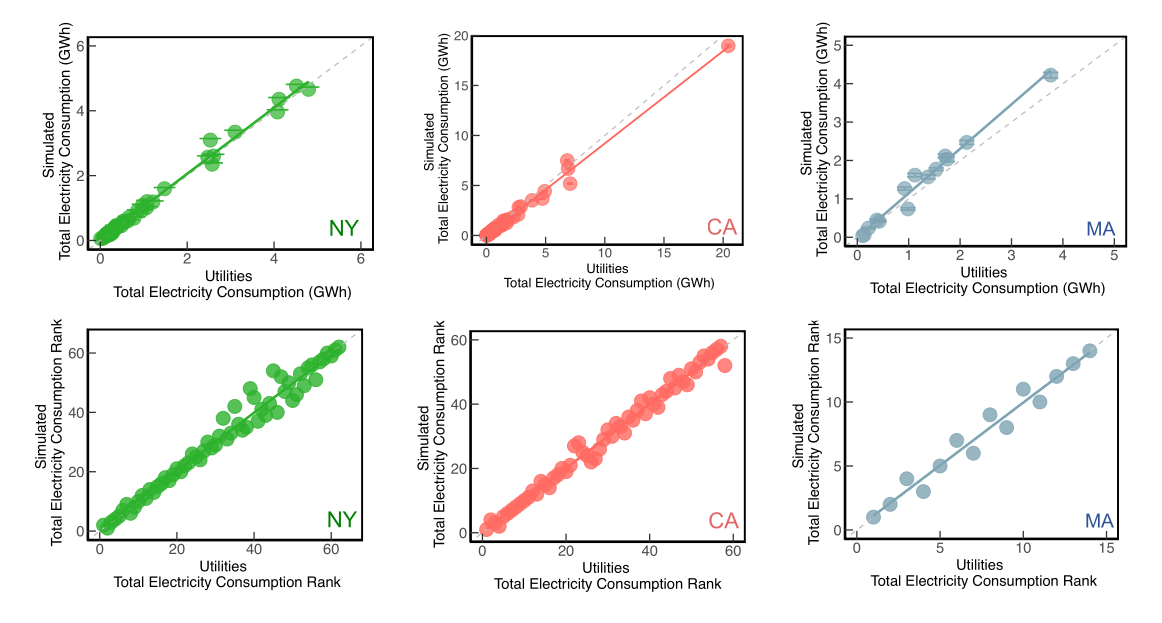}
\caption{}
\label{fig:fig_6_2}
\end{subfigure}
\end{figure}
\begin{figure}[ht]
\ContinuedFloat
\centering
\begin{subfigure}[b]{0.7\textwidth}
\includegraphics[width=\linewidth]{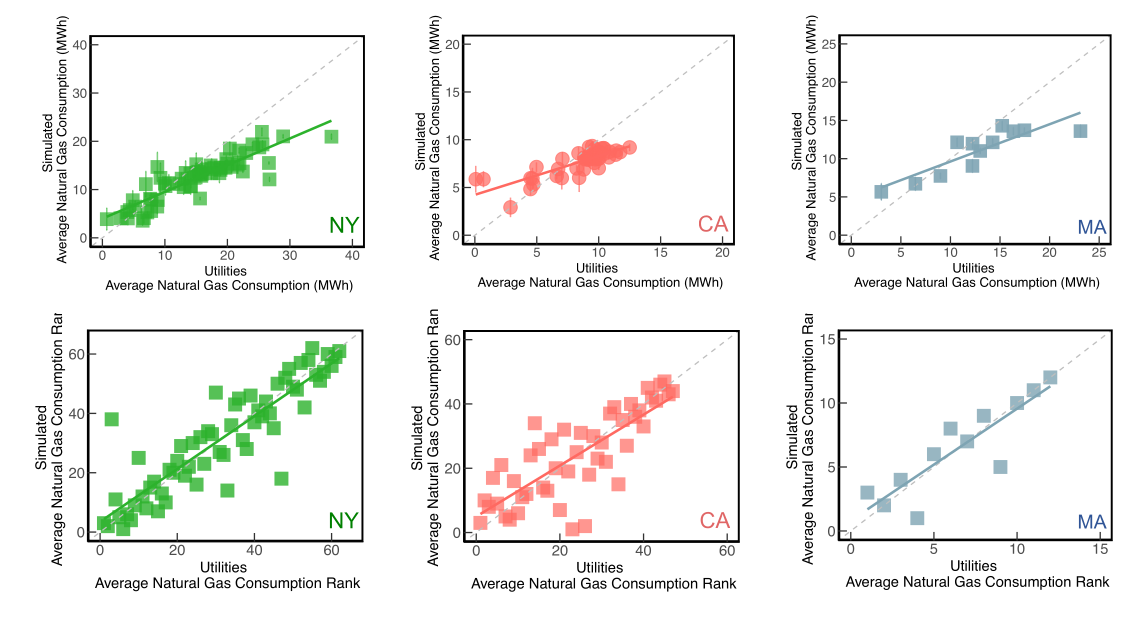}
\caption{}
\label{fig:fig_6_3}
\end{subfigure}
\begin{subfigure}[b]{0.7\textwidth}
\includegraphics[width=\linewidth]{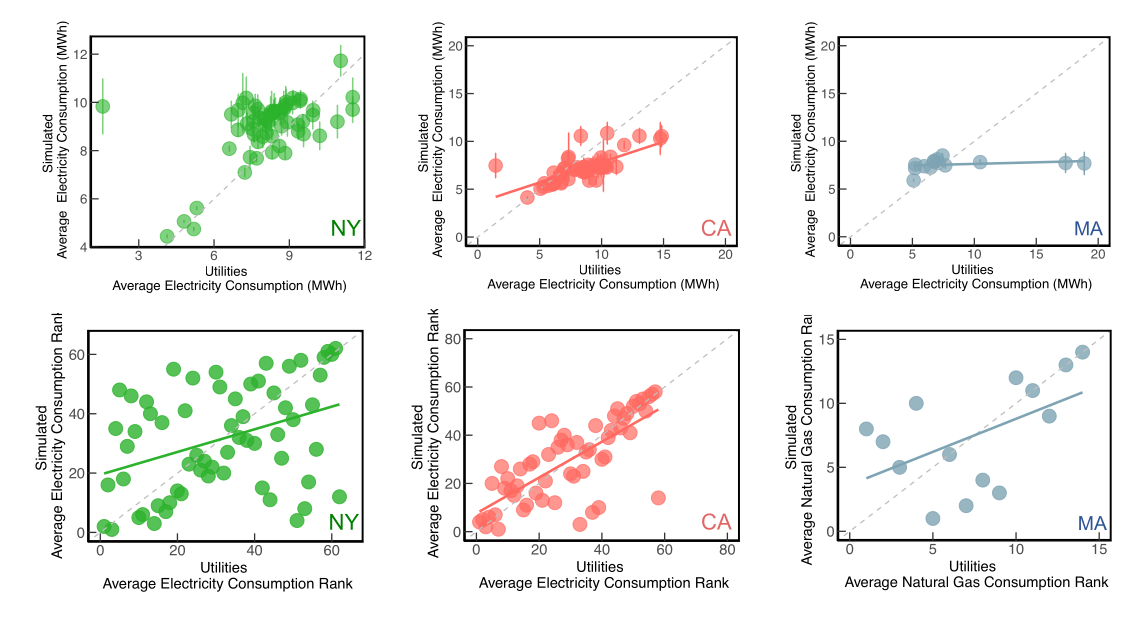}
\caption{}
\label{fig:fig_6_4}
\end{subfigure}
\begin{subfigure}[b]{0.7\textwidth}
\includegraphics[width=\linewidth]{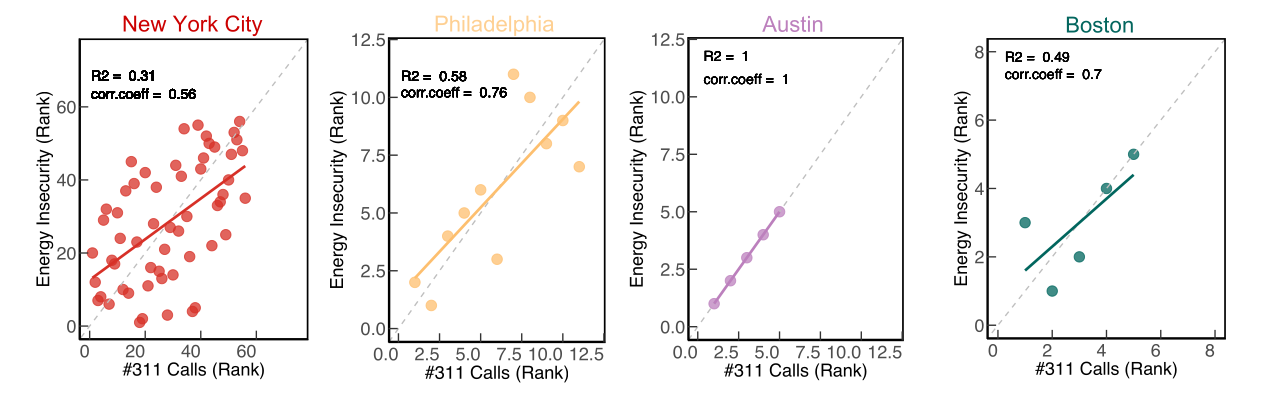}
\caption{}
\label{fig:fig_6_5}
\end{subfigure}
\caption{External validation of fusionACS using ACS-RECS fused dataset. Comparison of total \{(\ref{fig:fig_6_1}) simulated electricity and (\ref{fig:fig_6_2})\} simulated natural gas consumption metrics versus the actual total consumption metrics obtained from external sources, Comparison of average \{(\ref{fig:fig_6_3}) simulated electricity and (\ref{fig:fig_6_4})\} simulated natural gas consumption metric versus the actual consumption metrics obtained from external sources for counties in New York, California, and Massachusetts. Comparison of the rank of energy insecurity variable versus the rank of the number of 311 calls made at the PUMA-level for New York City, Philadelphia, Austin, and Boston\{(\ref{fig:fig_6_5})\}.}
\label{fig:fig_6}
\end{figure}

\begin{table}[ht]
\centering
\begin{tabular}{lll}
\textbf{Dataset} & \textbf{Resolution} & \textbf{Description} \\
\href{https://www.census.gov/programs-surveys/acs/microdata.html}{ACS-PUMS} & PUMA & PUMA-level summary values of ACS-PUMS variables. \\
\href{https://www.census.gov/programs-surveys/acs/data/summary-file.html}{ACS-SF} & block & Additional sociodemographic variables solicited by ACS. \\
\href{https://www.eia.gov/state/seds/}{EIA-SEDS} & state & Residential fuel prices and average consumption. \\
\href{https://www.epa.gov/smartgrowth/smart-location-mapping}{EPA-SLD} & block group & Variables describing the built environment, transit, walkability, etc. \\
\href{https://www.irs.gov/statistics/soi-tax-stats-statistics-of-income}{IRS-SOI} & zip code & Summary of information reported on 1040 personal tax returns. \\
\href{https://apps.openei.org/USURDB/}{NREL-URDB} & zip code & Residential average electricity prices.
\end{tabular}
\caption{Spatial Datasets Used}
\label{tab:tab_1}
\end{table}

\begin{table}[ht]
\centering
\small
\begin{tabular}{lll}
\textbf{Dataset} & \textbf{Variable Name} & \textbf{Variable Description} \\
\hline
\hline
RECS 2015 & btung & Household natural gas usage, in btu \\
& btulp & Household propane usage, in btu \\
&btufo & Household oil/kerosene usage, in btu \\
& btuel & Household electricity usage, in btu \\
& cooltype & Type of air conditioning equipment used \\
& insec & Faced some form of energy insecurity in the last year \footnote{Constructed from “scalee”, ”scaleg”, ”scaleb”} \\
& noac & In last year, was the household ever unable to use A/C because it could not \\
& & afford electricity or equipment repair? \footnote{Constructed from “noacel” and “noacbroke”.} \\
& noheat & In the last year, was the household ever unable to use heating equipment \\
& & because it could not afford energy or equipment repair? \footnote{Constructed from “noheatbroke”, “noheatbulk”,  “noheatel”, and “noheatng”.} \\
& dollarng & Household natural gas expenditure, in US dollars \\
& dollarlp & Household propane expenditure, in US dollars \\
& dollarfo & Household oil/kerosene expenditure, in US dollars \\
& dollarel & Household electricity expenditure, in US dollars \\
\hline
AHS 2019 &  cold & Flag indicating unit was uncomfortably cold for 24 hours or more last winter \\
& hmreneff & Flag indicating home improvements done to make home more energy efficient \\
&  & in last two years \\
& hotwater & Type of hot water system \\
& ratinghs & Rating of unit as a place to live \\
& fsstatus & Rating of overall food security of the household \\
& unitsize & Unit size (square feet) \\
\hline
NHTS 2017 &  place & Travel is a financial burden \\
& price & Price of gasoline affects travel \\
& ptrans & Public transportation to reduce financial burden of travel \\
& travel & Walk/bike to reduce financial burden of travel \footnote{Constructed from “bike2save” and “walk2save”.} \\
& gstotcst & Annual fuel expenditures in US dollars \\
\hline
CEI 2015 & cloftw & Expenditure on clothing and footwear \\
 & jwlbg & Expenditure on jewelry and handbags \\
 & educ & Expenditure on education services \\
 & stdint & Student loan interest payments \\
 & eltrnp & Expenditure on electronic products \\
 & hotel & Expenditure on hotels and motels \\
 & oeprd & Expenditure on other entertainment products \\
 & oesrv & Expenditure on other entertainment services \\
 & recrp & Expenditure on recreational products \\
 & eathome & Expenditure on eating and drinking at home \\
 & eatout & Expenditure on eating and drinking out \\
 & health & Expenditure on health care and insurance premiums \\
 & furhwr & Expenditure on furniture, housewares, and tools \\
& happl & Expenditure on household appliances \\
 & hhpcp & Expenditure on household and personal care products \\
 & hhpcs & Expenditure on household, personal, and child care services \\
 & hinsp & Expenditure on home insurance, primary \\
 & hmtimp & Expenditure on home maintenance and improvement 
\end{tabular}
\caption{List of fused variables for the different datasets (continued on next page)}
\end{table}

\setcounter{table}{1}

\begin{table}[ht]
\centering
\begin{tabular}{lll}
\textbf{Dataset} & \textbf{Variable Name} & \textbf{Variable Description} \\
\hline
\hline
CEI 2015 & mrtgip & Mortgage interest payments, primary \\
 & mrtgpp & Mortgage principal payments, primary \\
 & mrtgps & Mortgage principal payments, secondary \\
 & ohouse & Other housing expenses \\
 & ptaxp & Property taxes, primary \\
 & rent & Rent \\
 & chrty & Charitable contributions \\
 & finpay & Insurance, financial services, and other payments \\
 & ocash & Other cash transfers \\
 & othint & Interest and finance charges on credit card and other debt \\
 & rntval & Annual rental value, imputed for owner-occupied homes \\
 & tax & Net tax burden (before-tax income minus after-tax income) \\
 & airshp & Expenditure on air and ship travel \\
 & gas & Expenditure on gasoline and other motor fuel \\
 & pubtrn & Expenditure on public transportation \\
 & taxis & Expenditure on taxi and ride sharing services \\
 & vehins & Vehicle insurance \\
 & vehint & Vehicle loan interest payments \\
 & vehmlr & Expenditure on vehicle maintenance, leasing, and rental \\
 & vehnew & Gross value of new vehicle purchases \\
 & vehprd & Expenditure on vehicle parts, accessories, and supplies \\
 & vehprn & Vehicle loan principal payments \\
 & vehreg & Vehicle licensing, registration, and inspection \\
 & vehval & Value of owned vehicles \\
 & elec & Expenditure on electricity \\
 & intphn & Expenditure on internet and phone \\
 & ngas & Expenditure on natural gas \\
 & ofuel & Expenditure on heating oil, LPG, and other fuels \\
 & watrsh & Expenditure on water, sewer, and trash 
\end{tabular}
\caption{List of fused variables for the different datasets (continued from previous page)}
\label{tab:tab_2}
\end{table}

\end{document}